# New Results from RENO and The 5 MeV Excess

Seon-Hee Seo[1, a)] for the RENO Collaboration

[1] *Seoul National University*
*Department of Physics & Astronomy*
*1 Gwanak-ro, Gwanak-gu, Seoul, 151-747, Korea*

[a)] Corresponding author: shseo@phya.snu.ac.kr

**Abstract.** One of the main goals of RENO (Reactor Experiment for Neutrino Oscillation) is to measure the smallest neutrino mixing angle $\theta_{13}$ using reactor neutrinos in Korea. RENO is the first reactor experiment taking data with two identical detectors in different locations (Near and Far), which is critical to reduce systematic uncertainty in reactor neutrino flux. Our data taking has been almost continuous since Aug. 2011 and we have collected about 434,000 (54,000) electron anti-neutrinos in the Near (Far) detector by 2013. Using this data (about 800 live days) we present a new result on $\theta_{13}$: $\sin^2 2\theta_{13} = 0.101 \pm 0.008$ (stat.) $\pm 0.010$ (syst.). We also report the 5 MeV excess present in the prompt signal spectrum in our data, and its correlation with our reactor thermal power.

## INTRODUCTION

RENO has measured the last unknown and the smallest neutrino mixing angle $\theta_{13}$ with 4.9 $\sigma$ significance in 2012 using 220 live days of data taken with both the Near and the Far detectors [1]. In 2013 RENO updated the measurement twice using increased statistics (403 live days) and improved systematics, resulting in better precisions (5.6 $\sigma$ [2], 6.3 $\sigma$ [3]). Precise measurements of neutrino mixing parameters are important for model building as well as allowing less precisely measured neutrino mixing parameters to be better determined.

In the Neutrino 2014 conference we present our updated result on the $\theta_{13}$ measurement with better precision using double the data set (about 800 live days) and improved systematic errors. We also report, in our data, an unexpected Inverse Beta Decay (IBD) event excess around the 5 MeV region of the prompt signal spectrum.

## THE RENO EXPERIMET

To measure the small mixing angle it is important to precisely measure the small deficit of neutrinos at the distance where the maximum deficit is expected, between 1 and 2 km. However, our current knowledge of absolute reactor neutrino flux is not precise enough (5 % uncertainty [4]) to determine the small deviation from the expected flux caused by oscillations. To achieve this we need two identical detectors: one in near and the other in far site. Using neutrinos observed in the near detector site one can precisely predict the flux of neutrinos at the far site as the distance to the far site is very accurately measured. The RENO collaboration built two almost identical detectors and has taken data continuously since August 2011 using the two detectors.

In this section I will briefly describe our experimental setup, the RENO detector, and our energy scale calibration in sequence. Further details on the RENO experiment can be found in [5].

## The Experimental Setup

The RENO detector detects electron anti neutrinos coming from Hanbit nuclear reactors in Yonggwang, the southwest part of Korea. There are a total of 6 reactor cores aligned in equidistance along a total 1280 m length. The total maximum thermal power from the 6 reactor cores is about 16.8 $GW_{th}$ corresponding to about $3 \times 10^{21}$ electron antineutrinos per second. The Near (Far) detector is positioned in 290 m (1380 m) distance perpendicular to the center of the 6-reactor-array. The Near (Far) detector is located underneath a tunnel of a small hill equivalent to an overburden of 110 (450) meter water equivalent (m.w.e.).

## The RENO Detector

The RENO detector consists of four layers of cylinders that contain different liquids to serve different purposes. The innermost layer is neutrino Target filled with 16 ton of liquid scintillator (Linear Alkyl Benzen [6]) with 0.11 % Gadolinium (Gd) loaded in an acrylic vessel (Radius = 1.4 m, Height = 3.2 m). The Target is enclosed by Gamma-catcher filled with 30 ton of unloaded liquid scintillator in an acrylic vessel (Radius = 2.0 m, Height = 4.4 m). The Gamma-catcher serves to increase fiducial volume by catching gammas that occur outside the Target region. The Gamma-catcher is enclosed by Buffer filled with 64 ton of mineral oil in a stainless steel vessel (Radius = 2.7 m, Height = 5.8 m) where a total of 354 PMTs (Hamamatsu R7081, 10 inch) were attached pointing inward. The Buffer is enclosed by the outermost Veto filled with 353 ton of purified water in a concrete tank (Radius = 4.2 m, Height = 8.8 m) where a total of 64 waterproof PMTs of the same type used in the Buffer were attached. The inner three layers of detector parts are called an inner detector (ID) and the veto detector is called an outer detector (OD). **FIGURE 1** illustrates the RENO detector components.

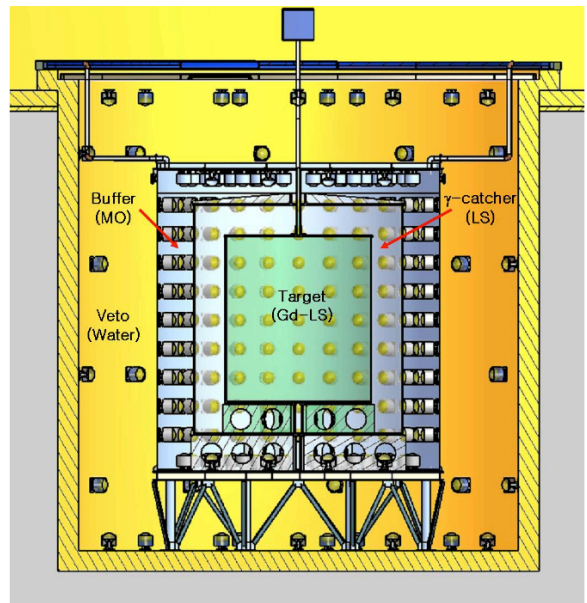

**FIGURE 1.** The layout of the RENO detector (side view).

## Energy Scale Calibration

To calibrate energy we use four radioactive sources with known energies: $^{137}$Cs (0.662 MeV), $^{68}$Ge (1.022 MeV), $^{60}$Co (2.5057 MeV), and $^{252}$Cf (n-H: 2.223 MeV, n-Gd: 7.937 MeV). To find a relation between photoelectron (p.e.) and energy, we use $^{68}$Ge, $^{60}$Co and IBD delayed signals (n-H and n-Gd) in our physics analysis data. **FIGURE 2** in the top panel shows p.e./energy(MeV) as a function of positron energy (MeV) where four dots with error bars are data points and the solid line represents a fitting line described as Eq. (1). The plots in the bottom panel of **FIGURE 2** show the fitting accuracies which are within 1 % for both the Near and the Far detectors.

$$y = A - \frac{B}{1 - \exp(-Cx - D)} \tag{1}$$

where y is a variable for energy (in MeV) and x is for p.e.

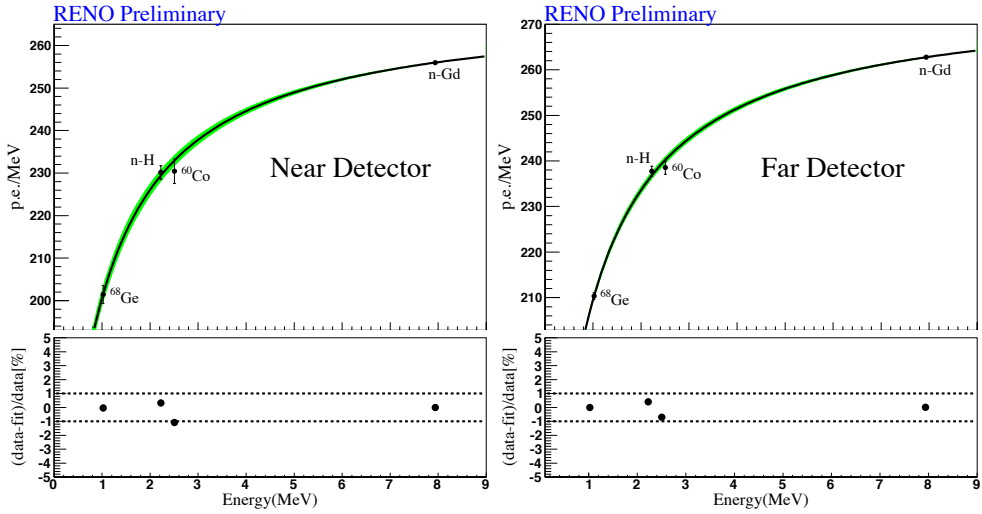

**FIGURE 2.** Top: p.e./MeV vs. positron energy. Bottom: Fitting accuracy vs. positron energy.

## DATA ANALYSIS

In the following subsections I will discuss data used for this analysis, IBD signal and background, event selection, and systematic errors in sequence.

### Data

RENO has been taking data continuously with both the Near and the Far detectors for more than 3 years since August 2011. The average DAQ efficiency is about 95 % (97 %) for the Near (Far) detector. In this analysis we used data taken from August 2011 to December 2013, corresponding to 761.11 and 794.72 live days of data for the Near and the Far detectors, respectively.

# IBD Signal and Background

When an electron anti neutrino hits a proton target in our detector, an IBD process ($\bar{\nu}_e + p \rightarrow e^+ + n$) could occur if the energy of the neutrino is greater than 1.8 MeV threshold required for an IBD process. Almost all kinetic energy of the neutrino goes to the positron and O(keV) goes to the neutron. The positron annihilates immediately and produces photons giving the neutrino energy spectrum. This immediate photon signal is called a prompt signal (or s1). The neutron, however, is captured by Gd (or Hydrogen) with a mean delay time of 30 μs in 0.11 % concentration of Gd (or 200 μs) and releases energy of 7.96 MeV (or 2.223 MeV), which can be used as a delayed signal (or s2). By requiring both s1 and s2 within a certain delay time, many backgrounds are suppressed. In this analysis we select the IBD events with neutron captured by Gd (n-Gd) only.

There are four types of backgrounds: accidental, fast neutron, $^9$Li/$^8$He and $^{252}$Cf backgrounds. These backgrounds still survive the final selection described in the following subsection. To measure the small $q_{13}$ value it is very important remove the remaining background very well.

Accidental background is formed when radioactivity gammas from our detector or surrounding environment mimic prompt signal and a thermal neutron (originally a fast neutron induced by atmospheric muon) captured by Gd. This uncorrelated prompt and delayed pair follows a Poisson statistics, and therefore accidental background was estimated using the following relation:

$$N_{accidental} = N_{s2} \times \left(1 - \exp^{[-R_{s1}(Hz) \times \Delta T(s)]}\right) \pm \frac{N_{accidental}}{\sqrt{N_{s2}}} \tag{2}$$

where, s1 and s2 events are counted just before IBD pairing (but after muon removal) by only requiring their corresponding energy range, i.e., [0.75, 12] MeV for s1 and [6, 12] MeV for s2, $R_{s1}$(Hz) is s1 event rate in Hz, and ΔT(s) is a coincidence time window in second, i.e., [2, 100] μsec between s1 and s2. From Eq. (2) the estimated accidental background rate in [0.75, 8] MeV region is 1.82 ± 0.11 /day (0.36 ± 0.01 /day) for the Near (Far) detector.

Fast neutron background fakes IBD signal by recoiling proton mimicking prompt signal and then being captured by Gd. Fast neutrons were estimated in an IBD signal search process with s1 energy spectrum extended up to 45~50 MeV. Above 18 MeV the s1 spectrum caused by fast neutrons appears flat and thus a fitting with a flat function was performed in [18~22, 45~47] MeV regions to estimate fast neutron background in the signal region where the same flat spectrum is assumed. The estimated fast neutron background rate in [0.75, 8] MeV signal region is 2.09 ± 0.06 /day (0.44 ± 0.02 /day) for the Near (Far) detector. The systematic error was obtained using alternative spectral shape, i.e., linear function instead of flat, and take 50 % of the difference.

$^9$Li and $^8$He are unstable isotopes and thus decay by emitting (β, n) followers. They are produced by atmospheric muons smashing $^{12}$C in liquid scintillator. The (β, n) followers fake IBD signals. The spectral shape of the $^9$Li/$^8$He background was obtained from event time distribution since last muon (> 1.5 GeV) where IBD and other background components are subtracted. The rate was estimated by "scaling" method. The "scaling" method estimates $^9$Li/$^8$He by obtaining a scale factor above 6.5 MeV between IBD candidates and pure $^9$Li/$^8$He sample in our data. This scale factor is used to estimate total $^9$Li/$^8$He background in [0.75, 8] MeV signal region by multiplying the scale factor to the pure $^9$Li/$^8$He sample in the same energy region. The estimated $^9$Li/$^8$He background rate in [0.75, 8] MeV signal region is 8.28 ± 0.66 /day (1.85 ± 0.20 /day) for the Near (Far) detector.

$^{252}$Cf background has existed since the end of October (September) 2012 for the Near (Far) detector due to a leakage of tiny fraction of $^{252}$Cf source, which was caused by loose O ring of the source container when taking $^{252}$Cf source data. It was found that the contamination is much less in the Near detector. The $^{252}$Cf background spectral shape and rate was obtained by subtracting all the other three background from IBD candidates for the Far detector. For the Near detector the $^{252}$Cf background spectral shape from the Far detector was used due to low statistics of

$^{252}$Cf events in the Near detector. And the rate was estimated by fitting IBD candidates with IBD signal and all background spectral functions by constraining with their own rates (except the $^{252}$Cf background) within their background systematic errors. The estimated $^9$Li/$^8$He background rate in [0.75, 8] MeV signal region is 0.28 ± 0.05 /day (1.98 ± 0.27 /day) for the Near (Far) detector.

## Event Selection

Our event selection criteria were based on maximizing signal to background ratio. To achieve this the following selection criteria were applied in sequence: (1) flasher and external gamma cut efficient for removing flasher events and accidental background, which is $Q_{max}/Q_{tot} < 0.03$, where $Q_{max}$ is the maximum charge deposited in a PMT among all hit PMTs in an event; (2) additional flasher cut to remove more active flasher events; (3) muon veto cuts (i) to rejects events within 1 ms following muons (with deposited energy ($E_m$) > 70 MeV, or 20 MeV < $E_m$ < 70 MeV with more than 50 hits in veto region); or (ii) to rejects events within 700 ms following muons ($E_m$ >1.5 GeV); (4) IBD signature cuts which accepts events with 0.7 MeV < $E(s1)$ < 12.0 MeV, 6.0 MeV < $E(s2)$ < 12.0 MeV, and [2, 100] μs time coincidence between s1 and s2; (5) several multiplicity cuts to remove fast neutron, $^9$Li/$^8$He, and $^{252}$Cf backgrounds.

After applying all selection criteria a total of 433,196 (50,750) IBD candidate events with $E_{s1}$ < 8 MeV were selected from the Near (Far) detector. After subtracting background described in the previous section the average daily observed IBD rates become 569.16 ± 0.87 /day and 63.86 ± 0.28 /day for the Near and the Far detectors, respectively, for the period of data we used in this analysis. **Figure 3** shows IBD prompt signal (solid line histogram) and backgrounds (hatched histograms) for the Near (left) and the Far (right) detectors. The inset plots are log-scale of the main plots to show better the small amount of backgrounds. The total background fraction is 3.1 % (8.1 %) for the Near (Far) detector. In **Figure 4** our daily observed and expected (with and without oscillation) IBD rates for each day of data taking are shown.

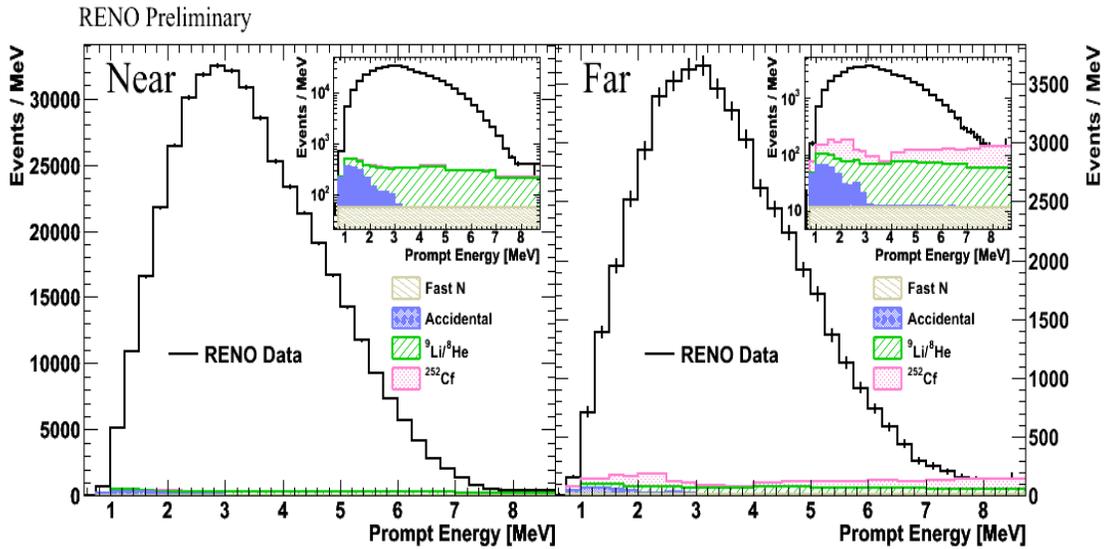

**FIGURE 3.** The IBD prompt signal spectrums for the Near (left) and the Far (right) detectors. The hatched histograms are backgrounds. The inset plots are the log scale (to show small background) plots of the main plots.

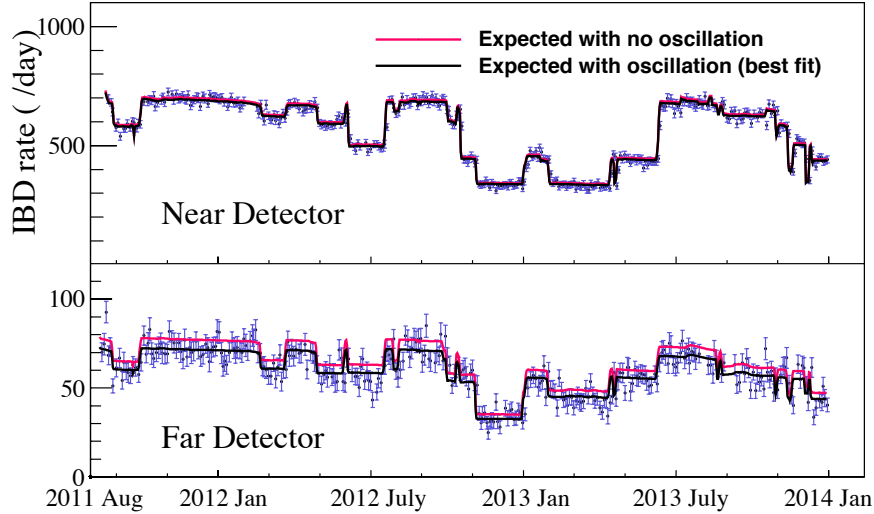

**FIGURE 4.** The daily IBD rates for the Near (top) and the Far (bottom) detectors. The dots with error bars represent observed IBD rates, and the solid lines are expected daily IBD rates without (in red) and with (in black) oscillation.

## Systematic Errors

The sources of the systematic errors of RENO come from three parts: reactor, detector and background. For the reactors the total systematic errors in our new result remain the same as our previous results [1-3] and they are 2.0 % and 0.9 % for correlated and uncorrelated errors, respectively.

For detectors the total correlated error was estimated as 1.29 % where the largest systematic error components are from Gd capture ratio (0.70 %) and "spill-in" systematic error (0.71 %). The total uncorrelated systematic error of our detectors was estimated as 0.20 %. This very small value of the total uncorrelated systematic error reflects the degree of identicalness of the two detectors, i.e., very identical.

For background our total systematic error was estimated as 5.4 (7.4) % for the Near (Far) detector as described in more detail in the previous subsection. The largest contribution to our background systematic error is from $^9$Li/$^8$He background (97 % contribution) for the Near detector and $^{252}$Cf and $^9$Li/$^8$He backgrounds (together 99 % contribution) for the Far detector. We still have a room to improve these background systematic errors.

## THE $\theta_{13}$ MEASUREMENT RESULTS

Using about 795 (761) live days of data taken in the Far (Near) we measured $\theta_{13}$ by minimizing $\chi^2$ with pull terms using rate-only analysis method used in [1-3]. **TABLE 1** summarizes the final data set used in the $\theta_{13}$ fitting. The total IBD rate after background subtraction is 569.16 ± 0.87 /day (63.86 ± 0.28 /day) for the Near (Far) detector. The total background rate is 12.48±0.68 /day (4.62 ± 0.28 /day) for the Near (Far) detector. The measured value of $\sin^2 2\theta_{13}$ is

$$\sin^2 2\theta_{13} = 0.101 \pm 0.008 \text{ (stat.)} \pm 0.010 \text{ (syst.)}.$$

**FIGURE 5** shows observed IBD rate (/day) versus expected IBD rate (/day) with oscillation using best-fit value of $\theta_{13}$. The fitted slope is $0.998 \pm 0.010$ and the fitted interception is $0.207 \pm 0.856$, which implies that our background is well subtracted and our data and expectation agrees very well. The spectral shape analysis on the same data set is also on going.

**TABLE 1.** Event rates of the observed IBD candidates and the estimated background.

| Detector | Near | Far |
|---|---|---|
| Selected events | 433,196 | 50,750 |
| Total background rate (per day) | 12.48±0.68 | 4.62± 0.28 |
| IBD rate after background subtraction (per day) | 569.16±0.87 | 63.86±0.28 |
| Live time (days) | 761.11 | 794.72 |
| Accidental rate (per day) | 1.82±0.11 | 0.36±0.01 |
| $^9$Li/$^8$He rate (per day) | 8.28±0.66 | 1.85±0.20 |
| Fast neutron rate (per day) | 2.09±0.06 | 0.44±0.02 |
| $^{252}$Cf rate (per day) | 0.28±0.05 | 1.98±0.27 |

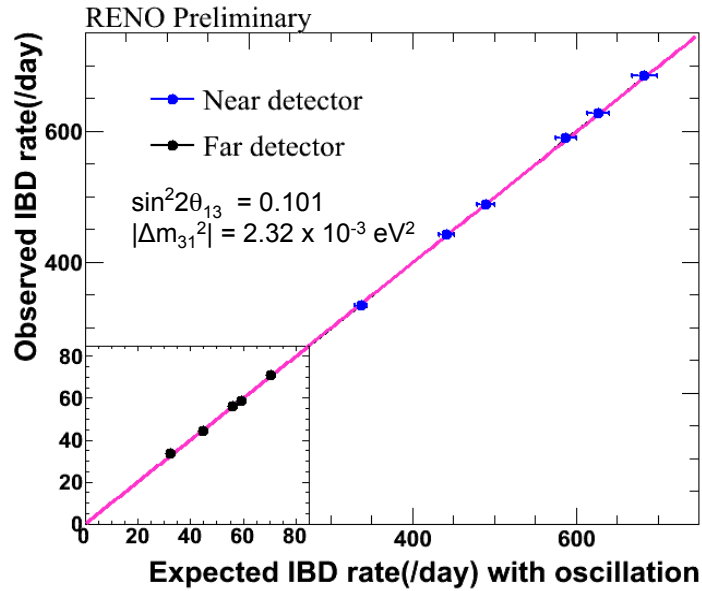

**FIGURE 5.** Observed IBD rate (/day) versus expected IBD rate (/day) for the Near (6 points from top) and the Far (5 points from bottom) detectors. The observed and expected rates match very well.

# THE 5 MEV EXCESS

As shown in **FIGURE 6** we observed excess of IBD events around 5 MeV with respect to the expected events obtained using reactor neutrino flux models by Mueller [7] and Huber [8]. The excess observed is:

Near: 2.303 % ± 0.401 % (experimental error) ± 0.492 % (expected error),

Far: 1.775 % ± 0.708 % (experimental error) ± 0.486 % (expected error)

where, the expected error is from the reactor neutrino flux models used in this analysis. The systematic error contribution to the experimental error above was estimated using energy scale, normalization, isotope fraction, MC modeling, spectral shape, and oscillation parameters uncertainties.

The excesses from the Near and the Far detectors are consistent within their errors. In the Near detector, the significance of excess is 3.5 $\sigma$. **FIGURE 7** shows correlation of the daily 5 MeV excess rate and the daily expected IBD rate (with oscillation using best fit) in the the Near detector, where the expected IBD rate was obtained taking into account our reactor thermal power and a daily fission fraction information. This correlation implies that the 5 MeV excess originates from reactor but not from known or unknown background that is independent with reactor thermal power.

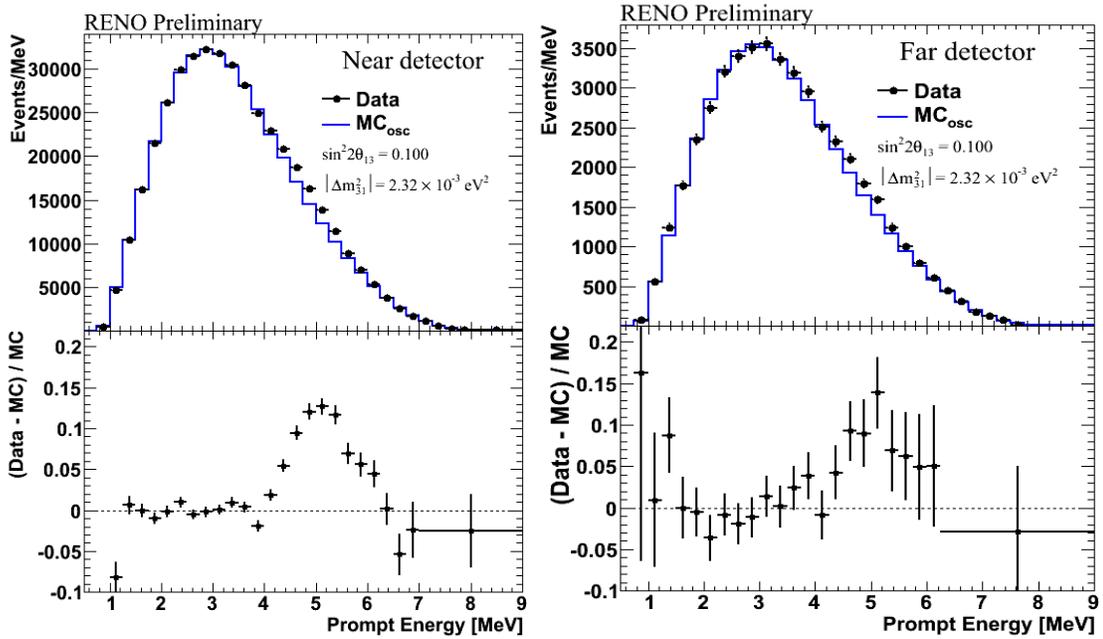

**FIGURE 6.** The IBD prompt signal spectrums of the observed (dots with error bars) and the expected events with oscillation (histogram) for the Near (top left) and the Far (top right) detectors. The bottom plots show the difference (data - expected) normalized to the expected events. The excesses of events around 5 MeV are observed in both the Near and the Far detectors.

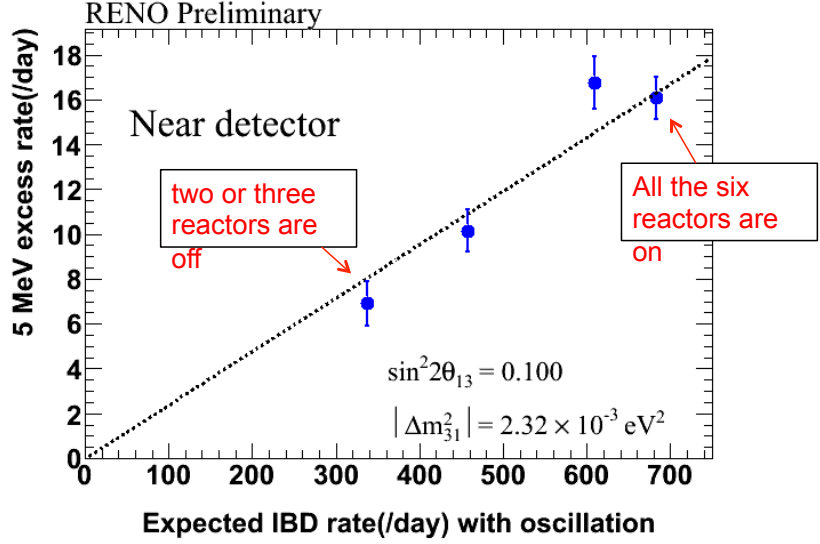

**FIGURE 7.** The correlation between the 5 MeV excess rate and the expected IBD rate with oscillation in the Near detector.

## CONCLUSIONS AND PROSPECTS

Using about 800 live days of data with rate-only analysis we measured $\sin^2 2\theta_{13} = 0.101 \pm 0.008$ (stat.) $\pm 0.010$ (syst.), corresponding to 7.8 $\sigma$. Spectral shape analysis is on-going for the same data set. We observed excess of IBD signal events at 5 MeV region compared to expectation based on the Mueller and Huber models [23, 24] of reactor neutrinos. The excess at the Near detector is 2.8 % with 3.5 $\sigma$ significance which takes into account the expected shape error from the models as well as our experimental errors. We showed that this excess has a correlation with reactor thermal power. This excludes background hypothesis for the excess. Our goal on the $\sin^2 2\theta_{13}$ value measurement is to reach 7 % precison by taking data for 5 years in total.

## ACKNOWLEDGMENTS

This work was supported by the Korea Neutrino Research Center which is established by the National Research Foundation of Korea (NRF) grant funded by the Korea government (MSIP) (No. 2009-0083526).